\documentclass[doublecol]{epl2}

\usepackage{amssymb}
\usepackage{amsmath}

\title{Extra current and integer quantum Hall conductance in the spin-orbit
coupling system}

\shorttitle{Extra current and IQH conductance in the SOC system}

\author{Y. Li \and T. Ma \footnote{Present address: Max Planck Institute for the Physics of
Complex Systems, D-01187 Dresden, Germany, EU.} \and R. B. Tao}

\shortauthor{Y. Li \etal}

\institute{Department of Physics, Fudan University - Shanghai
200433, China }



\abstract{ We study the extra term of particle current in a 2D
k-cubic Rashba spin-orbit coupling system and the integer
quantization of Hall conductance in this system. We provide a
correct formula of charge current in this system and the careful
consideration of extra currents provides a stronger theoretical
basis for the theory of quantum Hall effect which has not been
considered before. The nontrivial extra contribution to particle
current density and local conductivity, which originates from cubic
dependence on the momentum operator in the Hamiltonian, will have no
effect on the integer quantization of Hall conductance. The
extension of Noether's theorem for the 2D k-cubic Rashba system is
also addressed. Two methods reach to exactly the same results. }

\begin{document}

\maketitle


\section{Introduction}

Recent experimental demonstrations of spin Hall effect in some
semiconductors \cite{Kato, KatoPRL, Wunderlich, Sih2005, Silov} may
create a way to manipulate the spin of carriers in terms of electric
field that presents potential in future applications. It has
stimulated many scientists' interest. The experiments clearly show
that the spin-orbit coupling (SOC) of carriers in some
semiconductors plays a key role in disclosing the spin Hall effect.
Several theoretical models of SOC have been suggested to study the
charge and spin transport for different kinds of semiconductor
systems, such as $2$ dimensional ($2$D) linear $k$ dependent
Rashba\cite{linearRashba} and Dresselhaus\cite{linearDresselhaus1,
linearDresselhaus2} models, quadratic $k
$ dependent Luttinger model\cite{Luttinger}, $3$D\ $k$-cubic Dresselhaus model\cite%
{cubicDresselhaus}\ and $2$D $k$-cubic Rashba SOC model which was
found in a GaAs-Al$_{x}$Ga$_{1-x}$As interface of a typical
semiconductor heterojunction where the $k$-cubic Rashba SOC effect
for heavy holes can not be neglected in a high-density
regime\cite{cubicRashba}. For some Hamiltonians including terms with
high order power ($>2$) of momentum operators (MO)
$\widehat{\mathbf{p}}$, like $3$D $k$-cubic Dresselhaus model, we
have proved the conventional expression of particle current
density (CD) $\mathbf{j}_{conv}(\mathbf{r},t)=Re\{\psi ^{\dag }(\mathbf{r},t)%
1/(i\hbar)[\mathbf{r},\widehat{H}]\psi (\mathbf{r},t)\}=\left(
1/e\right) Re\{\psi ^{\dag }(\mathbf{r},t)(\partial
\widehat{H}/\partial \mathbf{A)}\psi (\mathbf{r},t)\}$ is no longer
valid\cite{us}. In that system,\ for the sake of current
conservation, a nontrivial extra term of CD
$\mathbf{j}_{extra}(\mathbf{r},t)$\ ($\nabla \cdot $\ $\mathbf{j}_{extra}(%
\mathbf{r},t)\neq 0$) should be added to the conventional one. Then
the continuity equation of conserved particle CD
$\mathbf{j}(\mathbf{r},t)\left(
=\mathbf{j}_{conv}(\mathbf{r},t)+\mathbf{j}_{extra}(\mathbf{r},t)\right)
$ can be satisfied.\ Thus, the extra term
$\mathbf{j}_{extra}(\mathbf{r},t)$\ is a physical quantity and has
effect on the conductivity of the system. It is natural to extend it
to $k$-cubic Rashba system where the extra terms of
CD may also appear due to cubic $k$, the term with high order power of MO $%
\widehat{\mathbf{p}}$\ in its Hamiltonian. However, $2$D $k$-cubic
Rashba is a real system that can demonstrate the integer quantum
Hall conductance. The high precision of integer quantum Hall
effect\cite{Klitzing80} was explained in some famous
papers\cite{Laughlin, TKNN, Kohmoto1985} where the expression of
charge CD is implicitly based upon the conventional form. One
naturally questions whether some correction to\ the quantum Hall
conductance could come from the additional term of current
$\mathbf{j}_{extra}(\mathbf{r},t)$\ in $2$D $k$-cubic Rashba SOC
semiconductors. In this paper, we would rigorously deduce the exact
expression of charge CD that shows the existence of nontrivial extra
term $\mathbf{j}_{extra}(\mathbf{r},t)$\ ($\nabla \cdot
\mathbf{j}_{extra}(\mathbf{r},t)\neq 0$). It is not a local circular
current and does have the contribution to electric conductivity.
Further, we prove that it has no contribution to the quantum Hall
conductance. So the explanation of integer quantum Hall conductance
is extended to a more general case that includes MO of triple power
in the Hamiltonian, though whose formula of charge CD must be
corrected by a nontrivial extra term due to the requirement of its
continuity. Our paper shows a more clear understanding of the
property of integer quantization of Hall conductance no matter the
Hamiltonian including additional cubic $k$ dependent SOC which is a
realizable $2$D quantum Hall system.

This paper is organized as following. Firstly, we simply introduce
the formulae of the calculation of particle CD in the first section.
The deduction of the
expression of extra term $\mathbf{j}_{extra}(\mathbf{r},t)$ in addition to $%
\mathbf{j}_{conv}(\mathbf{r},t)$ for a $2$D cubic Rashba Hamiltonian
is presented. The second section gives a proof that there is no
contribution to the
integer quantum Hall conductance from extra term $\mathbf{j}_{extra}(\mathbf{%
r},t)$. Our expression of particle CD confirmed by extended Noether's
theorem is attached in the appendix.

\section{Density of particle current}

We study the $2$D cubic Rashba system that is a promising model system for
an ultra thin film of $p$-doped semiconductor \cite{cubicRashba}. In a
perpendicular magnetic field, the single particle Hamiltonian is
\begin{eqnarray}
\widehat{H} &=&\widehat{H}_{N}\left( \widetilde{\mathbf{p}},\mathbf{r}%
\right) +\widehat{H}_{R},  \label{2} \\
\widehat{H}_{N}\left( \widetilde{\mathbf{p}},\mathbf{r}\right) &=&
\widehat{\widetilde{p}} ^{2}/2m^{\ast }
+V(\mathbf{r})-e\widehat{y}E_{y},  \label{3} \\
\widehat{H}_{R} &=&i\lambda (\widehat{\widetilde{p}}_{-}^{3}\sigma ^{+}-%
\widehat{\widetilde{p}}_{+}^{3}\sigma ^{-})  \label{4}
\end{eqnarray}%
where $V(\mathbf{r})$ is a local spin independent potential and
could contain an impurity potential, $\lambda =\alpha /2\hbar ^{3}$
is the spin-orbit
coupling constant, $E_{y}$ is the transverse Hall electric field, $\widehat{%
\widetilde{p}}_{\pm }=\widehat{\widetilde{p}}_{x}\pm i\widehat{\widetilde{p}}%
_{y},$ $\sigma ^{\pm }=\sigma _{x}\pm i\sigma _{y}$ where $\sigma _{x}$ and $%
\sigma _{y}$ are Pauli matrices, and $\widehat{\widetilde{\mathbf{p}}}%
=\left( \widehat{p}_{x}-eA_{x},\widehat{p}_{y}-eA_{y}\right) .$ The
corresponding Schr\"{o}dinger (or say Pauli) equation is%
\begin{equation}
\frac{\partial }{\partial t}\psi (\mathbf{r},t)=\frac{1}{i\hbar }\widehat{H}%
\psi (\mathbf{r},t),  \label{5}
\end{equation}%
where Hamiltonian $\widehat{H}$ is a $2\times 2$ matrix. The particle
density for a pure quantum state is $n(\mathbf{r},t)=\psi ^{\dag }(\mathbf{r}%
,t)\psi (\mathbf{r},t)$ in which we have performed the inner product for
spin space, but not for position. This rule of inner product is also used in
the following deductions implicitly. Since the number of particles is
conserved, the total number of particles $N=\int n(\mathbf{r},t)d\mathbf{r}$
should be a constant. The conserved particle CD $\mathbf{j}(\mathbf{r},t)$
is defined by the following continuity equation:
\begin{equation}
\frac{\partial n(\mathbf{r},t)}{\partial t}=-{\nabla \cdot \mathbf{j}(%
\mathbf{r},t).}  \label{PCD}
\end{equation}%
For simplifying the notations, in the paper, we will not discriminate the
notions of particle CD and charge CD which only differ by a factor of charge
$e$ and can be self-explanatory according to the context.

For a mixed state, the density matrix $\widehat{\rho }=\sum_{n}\left\vert
\psi _{n}\right\rangle \rho _{n}\left\langle \psi _{n}\right\vert $ where $%
\rho _{n}$ is the probability of the state $|\psi _{n}\rangle $, $\rho
_{n}\geq 0$, $\sum\limits_{n}\rho _{n}=1.$The density of particle is defined
by
\begin{eqnarray}
n(\mathbf{r},t) &=&\left\langle \mathbf{r,}t|\widehat{\rho }|\mathbf{r,}%
t\right\rangle =\sum_{n}\left\langle \mathbf{r,}t\left\vert \psi
_{n}\right\rangle \rho _{n}\left\langle \psi _{n}\right\vert \mathbf{r,}%
t\right\rangle  \notag \\
&\equiv &\sum_{n}\rho _{n}^{\dag }\psi _{n}(\mathbf{r},t)\psi _{n}(\mathbf{r,%
}t\mathbf{)},  \label{7}
\end{eqnarray}%
We discuss the case of $\rho _{n}$ being time independent. Then based on the
Schr\"{o}dinger equation, the left hand side of eq.(\ref{PCD}) can be
expressed as
\begin{eqnarray}
\frac{\partial n(\mathbf{r},t)}{\partial t}&=&\sum_{n}\rho
_{n}\{-\nabla \cdot
\mathbf{j}_{N}^{n}(\mathbf{r},t)  \notag\\
&&+(\frac{1}{i\hbar }\widehat{H}_{R}\psi _{n}(%
\mathbf{r},t))^{\dagger }\psi _{n}(\mathbf{r},t)  \notag\\
&&+\psi _{n}^{\dag }(\mathbf{r}%
,t)(\frac{1}{i\hbar }\widehat{H}_{R}\psi _{n}(\mathbf{r},t))\},
\label{continuity}
\end{eqnarray}%
where $\mathbf{j}_{N}^{n}(\mathbf{r},t)=Re\left\{ \psi _{n}{}^{\dagger }(%
\mathbf{r},t)\left(1/(i\hbar)[\mathbf{r},\widehat{H}_{N}]\psi _{n}{}(%
\mathbf{r},t)\right) \right\} $ is just the conventional expression
of
particle CD for non-SOC part $\widehat{H}_{N}\left( \widetilde{\mathbf{p}},%
\mathbf{r}\right) $. If the last two terms can be changed into $-\nabla
\cdot Re\left\{ \psi _{n}{}^{\dagger }(\mathbf{r},t)\left( 1/(i\hbar)[%
\mathbf{r},\widehat{H}_{R}]\psi _{n}{}(\mathbf{r},t)\right)
\right\}$, the particle CD would be
$\mathbf{j}^{n}(\mathbf{r},t)=Re\{ \psi
_{n}{}^{\dagger }(\mathbf{r},t)(1/(i\hbar)[\mathbf{r},\widehat{H%
}]\psi _{n}(\mathbf{r},t))\}$, which is just the conventional
formula $\mathbf{j}_{conv}^{n}(\mathbf{r},t)$. However, it is not
right. There should be an extra term
$\mathbf{j}_{extra}(\mathbf{r},t)$ in addition
to the term $\mathbf{j}_{conv}^{n}(\mathbf{r},t)$. And we will prove that $%
\nabla \cdot \mathbf{j}_{extra}(\mathbf{r},t)\neq 0.$ Thus, the conventional
formula of particle CD is not conserved in 2D cubic Rashba system. For
simplicity, we only consider the case in pure state and denote $\psi =\psi
_{n}(\mathbf{r},t)$. The expression of particle CD for mixed state can be
easily obtained from the one of pure state. After some algebra, we obtain%
\begin{eqnarray*}
&&\left( \frac{1}{i\hbar }\widehat{H}_{R}\psi \right) ^{\dagger }\psi +\psi
^{\dagger }\left( \frac{1}{i\hbar }\widehat{H}_{R}\psi \right) \\
&=&-\frac{1}{3}\nabla \cdot \left[ \left( \frac{1}{i\hbar }[\mathbf{r},%
\widehat{H}_{R}]\psi \right) ^{\dagger }\psi +\psi ^{\dagger }\left( \frac{1%
}{i\hbar }[\mathbf{r},\widehat{H}_{R}]\psi \right) \right] \\
&& +\frac{1}{3}%
\left( \frac{1}{i\hbar }[\mathbf{r},\widehat{H}_{R}]\psi \right) ^{\dagger
}\cdot \nabla \psi +\frac{1}{3}\left( \nabla \psi \right) ^{\dagger }\cdot \left( \frac{1}{%
i\hbar }[\mathbf{r},\widehat{H}_{R}]\psi \right) \\
&&-\frac{1}{3}\frac{eBy}{%
\hbar ^{2}}\left\{ \left( \left[ x,\widehat{H}_{R}\right] \psi \right)
^{\dagger }\psi +\psi ^{\dagger }\left[ x,\widehat{H}_{R}\right] \psi
\right\} \\
&=&-\nabla \cdot \mathbf{j}_{R}(\mathbf{r},t)\mathbf{.}
\end{eqnarray*}%
where%
\begin{eqnarray}
j_{R}^{x}(\mathbf{r},t) &=&\frac{1}{3}\left\{ \left( \frac{1}{i\hbar }[x,%
\widehat{H}_{R}]\psi \right) ^{\dagger }\psi +\psi ^{\dagger }\left( \frac{1%
}{i\hbar }[x,\widehat{H}_{R}]\psi \right) \right\} \notag \\
&&+\lambda Re\left\{ \left(
\partial _{y}\psi \right) ^{\dagger }\left[ x,\widetilde{p}_{-}^{2}\sigma
^{+}+\widetilde{p}_{+}^{2}\sigma ^{-}\right] \psi \right\}  \notag \\
&&+Re\left\{ \frac{\lambda eBy}{\hbar }\psi ^{\dagger }\left[ x,\widetilde{p}%
_{-}^{2}\sigma ^{+}-\widetilde{p}_{+}^{2}\sigma ^{-}\right] \psi
\right\}  \notag \\
&&-2\lambda \hbar ^{2}\left[ \left( \partial _{x}\psi \right)
^{\dagger }\sigma _{y}\partial _{x}\psi +\left(
\partial _{y}\psi \right) ^{\dagger
}\sigma _{y}\left( \partial _{y}\psi \right) \right]  \notag \\
&&-i2\lambda \hbar eBy\left[ \left( \partial _{y}\psi \right)
^{\dagger }\sigma _{x}\psi -\psi ^{\dagger }\sigma _{x}\partial
_{y}\psi \right]  \notag \\
&&+2\lambda \left( eB\right) ^{2}\left[ \psi ^{\dagger}y^{2}\sigma
_{y}\psi \right] , \label{jRx}
\end{eqnarray}%
\begin{eqnarray}
j_{R}^{y}(\mathbf{r},t) &=&\frac{1}{3}\left\{ \left( \frac{1}{i\hbar }[y,%
\widehat{H}_{R}]\psi \right) ^{\dagger }\psi +\psi ^{\dagger }\left( \frac{1%
}{i\hbar }[y,\widehat{H}_{R}]\psi \right) \right\}  \notag \\
&& +\lambda Re\left\{ \left(
\partial _{y}\psi \right) ^{\dagger }\left[ y,\widetilde{p}_{-}^{2}\sigma
^{+}+\widetilde{p}_{+}^{2}\sigma ^{-}\right] \psi \right\}  \notag \\
&&+Re\left\{ \frac{\lambda eBy}{\hbar }\psi ^{\dagger }\left[ y,\widetilde{p}%
_{-}^{2}\sigma ^{+}-\widetilde{p}_{+}^{2}\sigma ^{-}\right] \psi
\right\}  \notag \\
&&+2\lambda \hbar ^{2}\left[ \left( \partial _{x}\psi \right)
^{\dagger }\sigma _{x}\partial _{x}\psi +\left(
\partial _{y}\psi \right) ^{\dagger
}\sigma _{x}\partial _{y}\psi \right]  \notag \\
&&-i2\lambda \hbar eBy\left[ \left( \partial _{y}\psi \right)
^{\dagger }\sigma _{y}\psi -\psi ^{\dagger }\sigma _{y}\partial
_{y}\psi \right] \notag \\
&&-2\lambda \left( eB\right) ^{2}\left[ \psi ^{\dagger }y^{2}\sigma _{x}\psi %
\right] .  \label{jRy}
\end{eqnarray}%
In the above, following relations are applied:%
\begin{eqnarray*}
\widehat{H}_{R} &=&-\frac{1}{3}\nabla \cdot \lbrack \mathbf{r},\widehat{H}%
_{R}]+\frac{eBy}{3}\frac{1}{i\hbar }\left[ x,\widehat{H}_{R}\right] , \\
\widetilde{p}_{-}^{2}\sigma ^{+}+\widetilde{p}_{+}^{2}\sigma ^{-} &=&-\frac{1%
}{2}\nabla \cdot \left[ \mathbf{r},\widetilde{p}_{-}^{2}\sigma ^{+}+%
\widetilde{p}_{+}^{2}\sigma ^{-}\right] \\
&&+\frac{1}{2}eBy\frac{1}{i\hbar }%
\left[ x,\widetilde{p}_{-}^{2}\sigma ^{+}+\widetilde{p}_{+}^{2}\sigma ^{-}%
\right] .
\end{eqnarray*}%
So we have $\mathbf{j}(\mathbf{r},t)=\mathbf{j}_{N}(\mathbf{r},t)+\mathbf{j}%
_{R}(\mathbf{r},t)$. Then the extra term is
\begin{eqnarray*}
\mathbf{j}_{extra}(\mathbf{r},t) &=&\mathbf{j}(\mathbf{r},t)-\mathbf{j}%
_{conv}(\mathbf{r},t)  \\
&=&\mathbf{j}_{R}(\mathbf{r},t)-Re\left\{ \psi ^{\dagger }%
\frac{1}{i\hbar }[\mathbf{r},\widehat{H}_{R}]\psi \right\} ,
\end{eqnarray*}%
where $\mathbf{j}_{conv}(\mathbf{r},t)=Re\{ \psi ^{\dagger }
1/(i\hbar)[\mathbf{r},\widehat{H}]\psi \}$ is the so called
conventional current that widely appeared in literatures. Then the
expression of extra particle CD can be finally simplified as%
\begin{eqnarray}
j_{extra}^{x}(\mathbf{r},t) &=&2\lambda \hbar ^{2}\partial
_{x}\partial _{y}\left( \psi ^{\dagger }\sigma _{x}\psi \right)
-\lambda \hbar ^{2}\partial _{x}^{2}\left( \psi ^{\dagger }\sigma
_{y}\psi \right) \notag \\
&&+\lambda \hbar ^{2}\partial _{y}^{2}\left( \psi ^{\dagger }\sigma
_{y}\psi \right) ,
\label{extrax} \\
j_{extra}^{y}(\mathbf{r},t) &=&2\lambda \hbar ^{2}\partial
_{x}\partial _{y}\left( \psi ^{\dagger }\sigma _{y}\psi \right)
+\lambda \hbar ^{2}\partial _{x}^{2}\left( \psi ^{\dagger }\sigma
_{x}\psi \right) \notag \\
&&-\lambda \hbar ^{2}\partial _{y}^{2}\left( \psi ^{\dagger }\sigma
_{x}\psi \right) . \label{extray}
\end{eqnarray}%
In above equations, all $\left( \psi ^{\dagger }\sigma _{\alpha
}\psi \right) =(\psi ^{\dagger }(\mathbf{r},t),\sigma _{x}\psi
(\mathbf{r},t))$ are position dependent. The divergence of
$\mathbf{j}_{extra}(\mathbf{r},t)$ is generally non-zero, $ \nabla
\cdot \mathbf{j}_{extra}(\mathbf{r},t)=-\lambda \hbar ^{2}\partial
_{x}^{3}\left( \psi ^{\dagger}\sigma _{y}\psi \right) -\lambda \hbar
^{2}\partial _{y}^{3}\left( \psi ^{\dagger}\sigma _{x}\psi \right)
+3\hbar ^{2}\lambda \partial _{x}\partial _{y}^{2}\left( \psi
^{\dagger}\sigma _{y}\psi \right) +3\lambda \hbar ^{2}\partial
_{x}^{2}\partial _{y}\left( \psi ^{\dagger}\sigma _{x}\psi \right)
\neq 0$. So, as shown in Eqs.(\ref{extrax}) and (\ref{extray}), we
derived non-trivial extra terms in the expression of conserved
particle CD of a $2$D cubic Rashba Hamiltonian. The same result of
extra currents can also be obtained by taking account of the gauge
invariance based on Noether's theorem. Its detail is presented in
appendix.

When we consider a mixed state, the extra term of charge CD can be expressed
as
\begin{equation}
\mathbf{j}_{extra}(\mathbf{r},t)=e\sum_{n}\rho _{n}\mathbf{j}%
_{extra}^{\left( n\right) }(\mathbf{r},t).  \label{13}
\end{equation}%
In fact, our Hamiltonian $H(=H_{N}+H_{R})$ is time independent, so the
particle density $n$ and charge CD $\mathbf{j}$ are position dependent only.

\section{Hall conductance}

Now we study the charge CD along $x$ direction which is $j^{(x)}(\mathbf{r}%
)=j_{conv}^{(x)}(\mathbf{r})+j_{extra}^{(x)}(\mathbf{r})$. We take the
integral with respect to $y$ for $j^{(x)}(\mathbf{r})$ to get the charge
current
\begin{equation}
I^{(x)}=\int_{L_{y}}j_{conv}^{(x)}(\mathbf{r})dy+\int_{L_{y}}j_{extra}^{(x)}(%
\mathbf{r})dy,  \label{14}
\end{equation}%
where $L_{y}$ is the width of the system. And denote the length of
the system as $L_{x}.$ Finally, $L_{x}$ and $L_{y}$ can approach to
infinity if the system becomes macroscopic. $I^{(x)}$ should not be
position $x$ dependent because of the particle conservation. Then we
take the integral of $x$ for $I^{(x)}$ :
\begin{eqnarray}
I^{(x)}&=&\frac{1}{L_{x}}\int_{L_{x}}\int_{L_{y}}j_{conv}^{(x)}(\mathbf{r})d%
\mathbf{r+}\frac{1}{L_{x}}\int_{L_{x}}\int_{L_{y}}j_{extra}^{(x)}(\mathbf{r}%
)d\mathbf{r}  \notag \\
&=&\frac{1}{\Omega }\iint\limits_{\Omega }j_{conv}^{(x)}(%
\mathbf{r})L_{y}d\mathbf{r} + \frac{1}{\Omega }\iint\limits_{\Omega }j_{extra}^{(x)}(%
\mathbf{r})L_{y}d\mathbf{r}  \notag \\
&=&I_{conv}^{(x)}+I_{extra}^{(x)}.  \label{current}
\end{eqnarray}%
where $\Omega =L_{x}L_{y}$, $I_{conv}^{(x)}=\frac{1}{\Omega }\iint\limits_{\Omega }j_{conv}^{(x)}(%
\mathbf{r})L_{y}d\mathbf{r}$ and $I_{extra}^{(x)}=\frac{1}{\Omega }\iint\limits_{\Omega }j_{extra}^{(x)}(%
\mathbf{r})L_{y}d\mathbf{r}$. From Eq.(\ref{extrax}), the extra part
of the current is
\begin{eqnarray}
I_{extra}^{(x)} &=&e\lambda \hbar ^{2}L_{y}\sum_{n}\rho _{n}\{ \frac{1}{%
\Omega }\iint\limits_{\Omega }[2\partial _{x}\partial _{y}\left(
\Psi _{n}^{\dagger }(\mathbf{r)}\sigma _{x}\Psi
_{n}(\mathbf{r)}\right)  \notag\\
&&-\partial
_{x}^{2}\left( \Psi _{n}^{\dagger }(\mathbf{r)}\sigma _{y}\Psi _{n}(\mathbf{%
r)}\right) +\partial _{y}^{2}\left( \Psi _{n}^{\dagger }(\mathbf{r)}\sigma
_{y}\Psi _{n}(\mathbf{r)}\right) ]d\mathbf{r}\}  \notag \\
&=&-e\lambda L_{y}\sum_{n}\rho _{n}\frac{1}{\Omega }\iint\limits_{\Omega }d%
\mathbf{r}\{2\left( \widehat{p}_{x}\widehat{p}_{y}\Psi _{n}(\mathbf{r)}%
\right) ^{\dagger }\sigma _{x}\Psi _{n}(\mathbf{r)}  \notag \\
&&+2\Psi _{n}^{\dagger }(\mathbf{r)}\sigma
_{x}\widehat{p}_{x}\widehat{p}_{y}\Psi _{n}(\mathbf{r)}
-2\left( \widehat{p}_{x}\Psi _{n}(\mathbf{r)}\right) ^{\dagger }\sigma _{x}%
\widehat{p}_{y}\Psi _{n}(\mathbf{r)}
\notag \\
&&-2\left( \widehat{p}_{y}\Psi _{n}( \mathbf{r)}\right) ^{\dagger
}\sigma _{x}\widehat{p}_{x}\Psi _{n}(\mathbf{r)}-\left(
\widehat{p}_{x}^{2}\Psi _{n}(\mathbf{r)}\right) ^{\dagger }\sigma
_{y}\Psi _{n}(\mathbf{r)}
\notag \\
&&-\Psi _{n}^{\dagger }(\mathbf{r)}\sigma _{y}%
\widehat{p}_{x}^{2}\Psi _{n}(\mathbf{r)}+2\left( \widehat{p}_{x}\Psi _{n}(%
\mathbf{r)}\right) ^{\dagger }\sigma _{y}\widehat{p}_{x}\Psi _{n}(\mathbf{r)}  \notag \\
&&+\left( \widehat{p}_{y}^{2}\Psi _{n}(\mathbf{r)}\right) ^{\dagger
}\sigma_{y}\Psi _{n}(\mathbf{r)}+\Psi _{n}^{\dagger }(\mathbf{r)}\sigma _{y}%
\widehat{p}_{y}^{2}\Psi _{n}(\mathbf{r)}
\notag \\
&&-2\left( \widehat{p}_{y}\Psi _{n}(\mathbf{r)}
\right) ^{\dagger }\sigma _{y}\widehat{p}_{y}\Psi _{n}(\mathbf{r)}%
\}.  \label{Iextrax}
\end{eqnarray}%
The terms in right side of the above equation become the spatial
inner product after the integration of $\mathbf{r}$ over the whole
space of the system. Since the operators
$\{\widehat{p}_{x},\widehat{p}_{y}\}$ are hermitian, as an example,
we have
\begin{eqnarray*}
&&\frac{1}{\Omega }\iint\limits_{\Omega }d\mathbf{r}\{2\left( \widehat{p}_{x}%
\widehat{p}_{y}\Psi _{n}(\mathbf{r)}\right) ^{\dagger }\sigma _{x}\Psi _{n}(%
\mathbf{r)\}} \\
&=&\frac{1}{\Omega }\iint\limits_{\Omega }d\mathbf{r}%
\{2\left( \widehat{p}_{y}\Psi _{n}(\mathbf{r)}\right) ^{\dagger }\sigma _{x}%
\widehat{p}_{x}\Psi _{n}(\mathbf{r)\}} \\
&=&\frac{1}{\Omega }\iint\limits_{\Omega }d\mathbf{r}\{2\left( \widehat{p}%
_{x}\Psi _{n}(\mathbf{r)}\right) ^{\dagger }\sigma _{x}\widehat{p}_{y}\Psi
_{n}(\mathbf{r)\}} \\
&=&\frac{1}{\Omega }\iint\limits_{\Omega }d\mathbf{r}\{2\left( \Psi _{n}(%
\mathbf{r)}\right) ^{\dagger }\sigma _{x}\widehat{p}_{x}\widehat{p}_{y}\Psi
_{n}(\mathbf{r)\}.}
\end{eqnarray*}%
Considering above property for inner product of position space in equation (%
\ref{Iextrax}), we can easily obtain $I_{extra}^{(x)}=0.$ No contribution to
Hall conductance from extra term of charge CD is proved. Finally, we have
\begin{eqnarray*}
I^{(x)} &=&L_{y}\frac{1}{\Omega }\iint\limits_{\Omega }j_{conv}^{(x)}(%
\mathbf{r})d\mathbf{r} \\
&=&eL_{y}Re\sum_{n}\rho _{n}\frac{1}{\Omega }\iint\limits_{\Omega }\left\{
\Psi _{n}^{\dagger }(\mathbf{r)}\frac{1}{i\hbar }[\widehat{x},\widehat{H}%
]\Psi _{n}(\mathbf{r)}\right\} d\mathbf{r} \\
&=&eL_{y}Re\sum_{n}\rho _{n}\left\langle \Psi _{n}\right\vert \frac{1}{%
i\hbar }[\widehat{x},\widehat{H}_{0}]\left\vert \Psi _{n}\right\rangle , \\
H_{0} &=&\widehat{\widetilde{p}}^{2}/(2m^{\ast})+i\lambda (\widehat{%
\widetilde{p}}_{-}^{3}\sigma
^{+}-\widehat{\widetilde{p}}_{+}^{3}\sigma ^{-})
\end{eqnarray*}%
Thus, the quantum Hall conductance is only from the conventional term $%
\mathbf{j}_{conv}(\mathbf{r},t).$ For $H_{0}$, the cubic $2$D Rashba model
without transverse electric field, its Schr\"{o}dinger equation is
\begin{equation*}
H_{0}|\Psi _{n}^{(0)}\rangle =E_{n}^{(0)}|\Psi _{n}^{(0)}\rangle .
\end{equation*}%
It has been solved exactly \cite{txma2006},
\begin{eqnarray}
E_{n}^{(0)} &=&(n+1/2)\hbar \omega, n\leq 2,  \notag \\
E_{n,s}^{(0)} &=&\left[ (n-1)+s\sqrt{\gamma ^{2}n(n-1)(n-2)+\frac{9}{4}}%
\right] \hbar \omega, \notag \\
&& n\geq 3.  \label{eigenvalues}
\end{eqnarray}%
where $\omega =eB/(mc),s=\pm 1$ and $\gamma =4\lambda m^{\ast }\sqrt{%
2\hbar eB/c}.$ The eigen energies in traditional quantum Hall effect are
Landau levels separated by gaps. Now the \textquotedblleft Landau levels" of
a $2$D cubic Rashba model have some modification for $n\geq 3$, but they
still keep the essential feature of the gap separation. The corresponding
eigenfunctions are
\begin{eqnarray}
|\Psi _{n}^{(0)}\rangle &=&\left(
\begin{array}{c}
0 \\
\phi _{n}%
\end{array}%
\right) ,n\leq 2,  \notag \\
|\Psi _{n}^{(0)}\rangle &=&|\Psi _{n,s}^{(0)}\rangle =\left(
\begin{array}{c}
C_{ns1}\phi _{n-3} \\
C_{ns2}\phi _{n}%
\end{array}%
\right) ,n\geq 3.  \label{eigenfunctions}
\end{eqnarray}%
where $\{C_{ns1},C_{ns2}\}$ are normalized constants,%
\begin{eqnarray*}
C_{ns1} &=&\frac{ic_{n,s}}{\sqrt{c_{n,s}^{2}+1}},C_{ns2}=\frac{1}{\sqrt{%
c_{n,s}^{2}+1}}, \\
c_{n,s} &\equiv &\frac{1}{\gamma \sqrt{n(n-1)(n-2)}}  \\
&& \times \left( -\frac{3}{2}+s \sqrt{\gamma
^{2}n(n-1)(n-2)+\frac{9}{4}}\right) ,
\end{eqnarray*}%
and $\phi _{n}$ is the wave function of harmonic oscillation type.
Impurities may result in widening out the \textquotedblleft Landau levels".
The conventional velocity operator in position space is $\widehat{\mathbf{v}}%
=1/(i\hbar)[\mathbf{r},H]=1/(i\hbar)[\mathbf{r},H_{0}]$. It is
easy to have $\widehat{\mathbf{v}}(\mathbf{k})=1/\hbar \nabla _{%
\mathbf{k}}E^{(0)}(\mathbf{k})$\ in $\mathbf{k}$\ space$.$ Following Laughlin%
\cite{Laughlin} or Kohmoto's\cite{Kohmoto1985} deduction, the
integer quantization of quantum Hall conductance can be obtained.
Here we will present a different approach to reach to the conclusion
of integer quantum Hall conductance for such a specific 2D cubic
Rashba system.

Since the eigen energies and wave functions of Schr\"{o}dinger
equation in the second quantization representation can be found
exactly, we also calculate the Hall conductance in linear response
approximation and it shows excellent consistency with integer
quantization of Hall conductance. More specifically, the Hall
conductance $\sigma _{xy}$\ of this system can be written as $\sigma
_{xy}=\sum_{n,s}N_{n,s}\left( \sigma _{xy}\right) _{n,s}$, where
$N_{n,s}$
 is the number of particles occupying the $\left( n,s\right) $-th
 \textquotedblleft Landau level" (here the \textquotedblleft Landau
level" is marked by two index, $n$ indicating the energy level of the system
without SOC, $s$ indicating the energy level splitting due to SOC). And $%
\left( \sigma _{xy}\right) _{n,s}$ is the one particle's contribution from
the $\left( n,s\right) $-th \textquotedblleft Landau level", by linear
response theory,
\begin{eqnarray}
&&\left( \sigma _{xy}\right) _{n,s}= \sum_{\left( n^{\prime \prime
},s^{\prime \prime }\right) \neq
\left( n,s\right) }\notag \\
&& \times [\frac{\left\langle \Psi _{n,s}^{(0)}\right\vert
\widehat{j_{x}}\left\vert \Psi _{n^{\prime \prime },s^{\prime \prime
}}^{(0)}\right\rangle \left\langle \Psi _{n^{\prime \prime
},s^{\prime \prime }}^{(0)}\right\vert H^{\prime }\left\vert \Psi
_{n,s}^{(0)}\right\rangle }{E_{y}L_{x}L_{y}\left( E_{n,s}^{\left(
0\right) }-E_{n^{\prime \prime },s^{\prime \prime }}^{\left(
0\right) }\right) }+h.c.]. \label{sigmans}
\end{eqnarray}%
where $H^{\prime }=-e\widehat{y}E_{y}$ and the electric field is
uniform. Now we adopt the Landau gauge
$\widehat{\widetilde{p}}_{x}=\hbar
k_{x}-eBy/c,\widehat{\widetilde{p}}_{x}=\widehat{p}_{y}$,
and introduce the operator of bosonic quasi particles $a=\sqrt{\frac{c}{2\hbar eB%
}}\left( \widehat{\widetilde{p}}_{x}-i\widehat{\widetilde{p}}_{y}\right) $, $%
a^{\dagger }=\sqrt{\frac{c}{2\hbar eB}}\left( \widehat{\widetilde{p}}_{x}+i%
\widehat{\widetilde{p}}_{y}\right) $, $\left[ a,a^{\dagger }\right]
=1$. Then we get%
\begin{equation*}
\widehat{H}_{0}=\hbar \omega \left(
\begin{array}{cc}
a^{\dagger }a+\frac{1}{2} & i\gamma a^{3} \\
-i\gamma a^{\dagger 3} & a^{\dagger }a+\frac{1}{2}%
\end{array}%
\right) .
\end{equation*}%
Then%
\begin{eqnarray}
\widehat{j}_{x}&=&\frac{e}{i\hbar }\left[ x,\widehat{H}_{0}\right]
\notag \\
&=&e\omega \sqrt{\frac{\hbar c}{2eB}}\left( a+a^{\dagger }\right) +\frac{%
3ie\gamma }{m}\sqrt{\frac{\hbar eB}{2c}}\left(
\begin{array}{cc}
0 & a^{2} \\
-a^{\dagger 2} & 0%
\end{array}%
\right) .  \label{opjx}
\end{eqnarray}%
And%
\begin{equation}
\widehat{H^{\prime }}=-e\widehat{y}E_{y}=-eE_{y}\left( \hbar k_{x}-\sqrt{%
\frac{\hbar c}{2eB}}\left( a+a^{\dagger }\right) \right) \cdot I,
\label{opHprime}
\end{equation}%
where $I$ is a unit matrix. Then using Eqs.(\ref{eigenfunctions}),
(\ref{opjx}) and (\ref{opHprime})
for matrix elements $\langle \Psi _{n,\pm }^{(0)}|\widehat{%
j_{x}}|\Psi _{n+1,\pm }^{(0)}\rangle $ and $\langle\Psi _{n,\pm
}^{(0)}|H^{\prime }|\Psi _{n+1,\pm }^{(0)}\rangle$, the selection
rules will be found. And the summation
over states in Hall conductance in Eq.(\ref{sigmans}) can be simplified as%
\begin{eqnarray*}
&&\left( \sigma _{xy}\right) _{n,\pm } \\
&=&\frac{1}{EL_{x}L_{y}}[\frac{%
\left\langle \Psi _{n,\pm }^{(0)}\right\vert
\widehat{j_{x}}\left\vert \Psi _{n+1,+}^{(0)}\right\rangle
\left\langle \Psi _{n+1,+}^{(0)}\right\vert H^{\prime }\left\vert
\Psi _{n,\pm }^{(0)}\right\rangle }{E_{n,\pm }^{\left( 0\right)
}-E_{n+1,+}^{\left( 0\right) }}  \\
&&+\frac{\left\langle \Psi _{n,\pm }^{(0)}\right\vert
\widehat{j_{x}}\left\vert \Psi _{n-1,+}^{(0)}\right\rangle
\left\langle \Psi _{n-1,+}^{(0)}\right\vert H^{\prime }\left\vert
\Psi _{n,\pm }^{(0)}\right\rangle }{E_{n,\pm }^{\left(
0\right) }-E_{n-1,+}^{\left( 0\right) }} \\
&&+\frac{\left\langle \Psi _{n,\pm }^{(0)}\right\vert \widehat{j_{x}}%
\left\vert \Psi _{n+1,-}^{(0)}\right\rangle \left\langle \Psi
_{n+1,-}^{(0)}\right\vert H^{\prime }\left\vert \Psi _{n,\pm
}^{(0)}\right\rangle }{E_{n,\pm }^{\left( 0\right) }-E_{n+1,-}^{\left(
0\right) }}  \\
&&+\frac{\left\langle \Psi _{n,\pm }^{(0)}\right\vert \widehat{j_{x}%
}\left\vert \Psi _{n-1,-}^{(0)}\right\rangle \left\langle \Psi
_{n-1,-}^{(0)}\right\vert H^{\prime }\left\vert \Psi _{n,\pm
}^{(0)}\right\rangle }{E_{n,\pm }^{\left( 0\right) }-E_{n-1,-}^{\left(
0\right) }}+h.c.]
\end{eqnarray*}%
Then using eqs.(\ref{eigenvalues}) and (\ref{eigenfunctions}), all the
elements $\left\langle \Psi _{n,\pm }^{(0)}\right\vert \widehat{j_{x}}%
\left\vert \Psi _{n\pm 1,\pm }^{(0)}\right\rangle $\ and $\left\langle \Psi
_{n\pm 1,\pm }^{(0)}\right\vert H^{\prime }\left\vert \Psi _{n,\pm
}^{(0)}\right\rangle $\ can be calculated without difficulty. After long but
straight algebraic deduction, we can finally obtain%
\begin{equation}
\left( \sigma _{xy}\right) _{\widetilde{n},\pm }=-\frac{e^{2}}{h}\frac{\Phi
_{0}}{\Phi },  \label{25}
\end{equation}%
where $\Phi _{0}=BL_{x}L_{y},\Phi =hc/e$. By summing up all the
contributions from different energy levels, the total Hall conductance will
be%
\begin{eqnarray}
\sigma _{xy} &=&\sum_{\widetilde{n}}\left( \left( \sigma _{xy}\right) _{%
\widetilde{n},+}+\left( \sigma _{xy}\right) _{\widetilde{n},-}\right)
=-\sum_{\widetilde{n},s}N_{\widetilde{n},s}\frac{e^{2}}{h}\frac{1}{\Phi
/\Phi _{0}}  \notag \\
&=&-\frac{e^{2}}{h}\frac{N_{0}}{\Phi /\Phi _{0}}=-\nu \frac{e^{2}}{h}.
\label{26}
\end{eqnarray}%
Here $N_{0}$ is the total number of carriers, and the filling factor $\nu
=N_{0}/\left( \Phi /\Phi _{0}\right) $.\textsl{\ }

Due to the existence of impurities in practical samples, localized states
appear in the region between \textquotedblleft Landau levels". It leads to
the appearance of the plateaus when the Fermi level lies in that region. The
gap between two conductance plateaus is obviously $e^{2}/h$. It is concluded
that the cubic SOC do induce the extra term of CCD that yields the
contribution to electric conductivity, but no contribution is given to the
quantum Hall conductance.

\section{Conclusions}
We have derived an exact formula of particle current density for a
$2$D cubic Rashba model that appears in some $p$-doped
semiconductors. In addition to the conventional current expression,
there must be an extra term that ensures the current continuity
equation. The extra term must have the contribution to electric
conductivity, but no contribution to the charge quantum Hall
conductance that is proved rigorously. So, it can be clearly shown
that no effect is made on the topological property of integer
quantization of Hall conductance due to the existence of extra terms
in the $2$D cubic Rashba coupling system. Further experimentally
detectable effects of the new term are still on research.

\section{Appendix: Deduction of extra terms from Noether's theorem}

In this appendix, we point out that, for a $2$D cubic Rashba Hamiltonian
where the highest order of derivatives is higher than $2$, it is necessary
to generalize the expression of conserved current in Noether's theorem.
Applying the generalized Noether's theorem\cite{us}, we can get the
expressions of conserved particle CD of a $k$-cubic Rashba SOC system from $%
U\left( 1\right) $ gauge invariance.

Noether's theorem, not only indicates the relation between conserved
currents and symmetries of Lagrangian, but also implies that the
expression of conserved current depends on the form of Lagrangian
from the beginning
of its deduction. In usual cases, Lagrangians are expressed as $\mathcal{L}%
[\phi \left( x\right) ,\partial _{\mu }\phi \left( x\right) ,\phi
^{\dagger }\left( x\right) ,\partial _{\mu }\phi ^{\dagger }\left(
x\right) ],x^{\mu }=\left( t,\mathbf{r}\right) ,\mu =0,1,2,3$ -such
as the Lagrangian of complex scalar field- which only include fields
$\phi \left( x\right) ,\phi ^{\dagger }\left( x\right) $ and their
first order derivatives $\partial _{\mu }\phi \left( x\right)
,\partial _{\mu }\phi ^{\dagger }\left( x\right) $ as independent
variables. But in our case, Hamiltonian $\widehat{H}$ includes
higher order derivatives. So its Lagrangian should be written in the
form $\mathcal{L}[\phi (x) ,\partial _{\mu }\phi (x) ,\partial _{\mu
}\partial _{\nu }\phi (x) ,...,\phi ^{\dagger }(x) ,\partial _{\mu
}\phi ^{\dagger }(x) ,\partial _{\mu }\partial _{\nu }\phi ^{\dagger
}(x) ,...]$, where higher-order derivatives are also included as
independent variables. For simplicity, we denote $\phi \left(
x\right) $ and $\phi ^{\dag }\left( x\right) $ as $\phi $ and $\phi
^{\dag }$. The Hamiltonian of a $k$-cubic Rashba system studied here
is $\widehat{H}_{R}=\widehat{p}^{2}/\left(
2m\right) +i\lambda \left( \widehat{p}_{-}^{3}\sigma ^{+}-\widehat{p}%
_{+}^{3}\sigma ^{-}\right) $. The corresponding Lagrangian can be
\begin{eqnarray*}
&&\mathcal{L}[\phi ,\partial _{\mu }\phi ,\partial _{\mu }\partial _{\nu
}\phi ,\partial _{\mu }^{2}\partial _{\nu }\phi ;\phi ^{\dagger },\partial
_{\mu }\phi ^{\dagger },\partial _{\mu }\partial _{\nu }\phi ^{\dagger
},\partial _{\mu }^{2}\partial _{\nu }\phi ^{\dagger }] \\
&=&\phi ^{\dagger }\left( i\partial _{0}\phi \right)
+\frac{1}{2m}\phi ^{\dagger }\left( \partial _{x}^{2}+\partial
_{y}^{2}\right) \phi  \\
&& +2i\lambda \phi ^{\dagger }\left( \sigma ^{x}\partial
_{y}^{3}+\sigma ^{y}\partial _{x}^{3}\right) \phi \\
&&-2i\lambda \phi ^{\dagger }\left( 3\sigma ^{x}\partial
_{x}^{2}\partial _{y}+3\sigma ^{y}\partial _{x}\partial
_{y}^{2}\right) \phi.
\end{eqnarray*}%
According to the least action principle, one can easily obtain an
Euler-Lagrange equation
\begin{equation*}
0=\frac{\partial \mathcal{L}}{\partial \phi }-\partial _{\mu }\frac{\partial
\mathcal{L}}{\partial \left( \partial _{\mu }\phi \right) }+\partial _{\mu
}\partial _{\nu }\frac{\partial \mathcal{L}}{\partial \left( \partial _{\mu
}\partial _{\nu }\phi \right) }-\partial _{\mu }^{2}\partial _{\nu }\frac{%
\partial \mathcal{L}}{\partial \left( \partial _{\mu }^{2}\partial _{\nu
}\phi \right) },
\end{equation*}%
which yields the Schr\"{o}dinger equation. Actually, the first two
terms on the right-hand side of the above equation give the
conventional formula of particle CD.\ The remaining parts lead to
the extra terms. The corresponding conserved current $F_{\mu }$ is
\begin{align}
F^{\mu }=\frac{\partial \mathcal{L}}{\partial \left( \partial _{\mu
}\phi \right) }\delta \phi +\frac{\partial \mathcal{L}}{\partial
\left( \partial _{\mu }\partial _{\nu }\phi \right) }\delta \left(
\partial _{\nu }\phi \right)-(\partial _{\nu }\frac{\partial
\mathcal{L}}{\partial \left(
\partial _{\mu }\partial _{\nu }\phi \right) })\delta \phi \notag  \\
+\frac{\partial \mathcal{L}}{\partial \left( \partial _{\mu
}^{2}\partial _{\nu }\phi \right) }\delta \left( \partial _{\mu
}\partial _{\nu }\phi \right)-(\partial _{\mu }\frac{\partial
\mathcal{L}}{\partial \left( \partial _{\mu }^{2}\partial _{\nu
}\phi \right) })\delta \left( \partial _{\nu }\phi \right)  \notag
\\
+(\partial _{\nu }^{2}\frac{\partial \mathcal{L}}{\partial \left(
\partial _{\mu }\partial _{\nu }^{2}\phi \right) })\delta \phi +\left( \phi
\rightarrow \phi ^{\ast }\right),\tag{A.1}
\end{align}
which satisfies the continuity equation $\partial _{\mu }F^{\mu }
=0$. We concentrate on the deduction of conserved particle current
corresponding to $U\left( 1\right) $ gauge symmetry. From
infinitesimal variation of fields $\delta \phi =i\alpha \phi ,\delta
\phi ^{\dagger }=-i\alpha \phi ^{\dagger }$, the expression of
conserved particle CD for a $k$-cubic Rashba system is
\begin{eqnarray*}
\mathbf{j}^{x} &=&-F^{x}=\phi ^{\dagger }\left( \frac{i\partial
_{x}}{2m}\phi \right) +\left( \frac{i\partial _{x}}{2m}\phi \right)
^{\dagger }\phi
\\
&&-2\lambda [ \phi ^{\dagger }\sigma ^{y}\left( \partial
_{x}^{2}\phi \right) +\left( \partial _{x}^{2}\phi ^{\dagger
}\right) \sigma ^{y}\phi  \\
&&-\left( \partial _{x}\phi ^{\dagger }\right) \sigma ^{y}\left(
\partial _{x}\phi \right)]+6\lambda [ \phi ^{\dagger
}\left( \sigma ^{x}\partial _{x}\partial _{y}\phi \right) \\
&& -\left(
\partial _{x}\phi ^{\dagger }\right) \left( \sigma ^{x}\partial
_{y}\phi \right) +\left(
\partial _{y}^{2}\phi ^{\dagger
}\right) \left( \sigma ^{y}\phi \right)] , \\
\mathbf{j}^{y} &=&-F^{y}=\phi ^{\dagger }\left( \frac{i\partial
_{y}}{2m}\phi \right) +\left( \frac{i\partial _{y}}{2m}\phi \right)
^{\dagger }\phi \\
&&-2\lambda [ \phi ^{\dagger }\sigma ^{x}\left(
\partial _{y}^{2}\phi \right) +\left( \partial _{y}^{2}\phi
^{\dagger }\right) \sigma ^{x}\phi  \\
&&-\left( \partial _{y}\phi ^{\dagger }\right) \sigma ^{x}\left(
\partial _{y}\phi \right) ]+6\lambda [ \phi ^{\dagger }\left(
\sigma ^{y}\partial_{x}\partial _{y}\phi \right) \\
&& -\left( \partial _{y}\phi ^{\dagger }\right) \left( \sigma
^{y}\partial _{x}\phi \right) +\left(
\partial _{x}^{2}\phi ^{\dagger }\right) \left( \sigma ^{x}\phi
\right) ] .
\end{eqnarray*}%
Comparing the above formulae with the conventional one $\mathbf{j}%
_{conv}=Re\left\{ \phi ^{\dagger }\left( \frac{1}{i}\left[ \mathbf{r},H_{R}%
\right] \phi \right) \right\} $, we get the extra term of particle CD $%
\mathbf{j}_{extra}=\mathbf{j}-\mathbf{j}_{conv}$:%
\begin{align}
\mathbf{j}_{extra}^{x} &=&-\lambda \partial _{x}^{2}\left( \phi
^{\dagger }\sigma ^{y}\phi \right) +6\lambda \left( \partial
_{x}\phi ^{\dagger }\right) \sigma ^{x}\left( \partial _{y}\phi
\right)  \notag \\
&&+6\lambda \left(
\partial _{x}\partial _{y}\phi ^{\dagger }\right) \left( \sigma ^{x}\phi
\right) -6\lambda \left( \partial _{y}^{2}\phi ^{\dagger }\right) \left(
\sigma ^{y}\phi \right)  \notag \\
&&+3\lambda \phi ^{\dagger }\left( \sigma ^{y}\partial _{y}^{2}\phi \right)
+3\lambda \left( \partial _{y}^{2}\phi ^{\dagger }\right) \left( \sigma
^{y}\phi \right) ,  \tag{A.2} \\
\mathbf{j}_{extra}^{y} &=&-\lambda \partial _{y}^{2}\left( \phi
^{\dagger }\sigma ^{x}\phi \right) +6\lambda \left( \partial
_{y}\phi ^{\dagger }\right) \sigma ^{y}\left( \partial _{x}\phi
\right)  \notag \\
&&+6\lambda \left(
\partial _{x}\partial _{y}\phi ^{\dagger }\right) \left( \sigma ^{y}\phi
\right) -6\lambda \left( \partial _{x}^{2}\phi ^{\dagger }\right) \left(
\sigma ^{x}\phi \right)  \notag \\
&&+3\lambda \phi ^{\dagger }\left( \sigma ^{x}\partial _{x}^{2}\phi
\right) +3\lambda \left( \partial _{x}^{2}\phi ^{\dagger }\right)
\left( \sigma ^{x}\phi \right) .  \tag{A.3}
\end{align}%
Further, it is not difficult to check that the extra term $\mathbf{j}%
_{extra} $ deduced here by extended Noether's theorem and $\mathbf{j}%
_{extra} $ in the second section do satisfy the equation $\nabla \cdot \left( \mathbf{%
j}_{extra}-\mathbf{j}_{extra}\right) =0$. Thus we conclude that our result
of extra term is rigorous.

\acknowledgments This work is supported by the National Natural
Science Foundation of China (Nos. 10674027 and 10547001) and 973
project of China.

\end{document}